# Anti-sites disordering suppression of the possible phase transition in Mn$_2$CrGa

H. G. Zhang[1, 2], J. Chen[1], E. K. Liu[2], M. Yue[1]*, Q. M. Lu[1], W. H. Wang[2], G. H. Wu[2]

1) College of Materials Science and Engineering, Beijing University of Technology, Beijing 100124, China

2) State Key Laboratory of Magnetism, Institute of Physics, Chinese Academy of Sciences, Beijing 100190, China

Theoretical and experimental characterizations of Mn$_2$CrGa compound in regard to the possibility of phase transformation have been carried out in this work. Under a high ordering L2$_1$ structure, this compound has the potential to be a martensite phase transition material. However, experimental results show a severe disordering took place in this system, which forbids the occurring of the phase transition. This work provides important reference for the design of new phase transition materials in Heusler alloys.

1. INTRODUCTION

Mn-based Heusler alloy is one of the most abundant resources for ferromagnetic shape memory alloys, such as Mn$_2$NiZ[1,2], MnNiCoAl[3] and Mn$_2$PtIn[4], due to their martensite transformation from cubic phase to tetragonal phase. They are also promising candidates for the tunneling magnetoresistance, spin-transfer-torque-random-access memory applications and fundamental skyrmion related research.[5-8] Some works even suggest that those compounds stabilized in tetragonal phase with uniaxial anisotropy, large magnetic moment and high Curie temperature may suitable for rare-earth-free hard magnets.[9]

As an important member of this family, other than Mn$_2$(Fe, Co, Ni, Cu)Z, Mn$_2$CrZ has not yet been thoroughly investigated. Only few theoretical works suggested that they are all possible half-metallic ferro/anti-ferromagnets[10,11], without considering the possibility of phase transformation. First-principles calculation is one effective measure to explore new Mn-based Heusler alloys with phase transformation properties [11,12]. By comparing the ground state total energy of the cubic austenite and non-modulated tetragonal martensite phases, one can predict the possible phase transformation of a composition under certain circumstance[4,11,13-17]. However, deviation of the actual materials from the ideal structure is one big obstacle to the realization of phase transformation in these systems.[18] This is particularly true in Fe-based Heusler alloys. In $Fe_2Cu$(Al, Ga, Ge) and $Fe_2CoGe$ compounds, which were predicted to be tetragonal when being chemically ordered, the anti-sites disordering stabilize the system in cubic structure instead of the tetragonal one. [18] however, this point has not been discussed in Mn-based compounds.

In this work, we focus on the theoretical and experimental investigation of $Mn_2CrGa$ compound in regard to the possibility of phase transformation. This compound has been reported as a half-metallic antiferromagnets with magnetic moment of $1\mu_B$ per unit cell.[10] Our first-principles calculation, however, indicates that it maybe show a phase transition under at ideal ordering state.

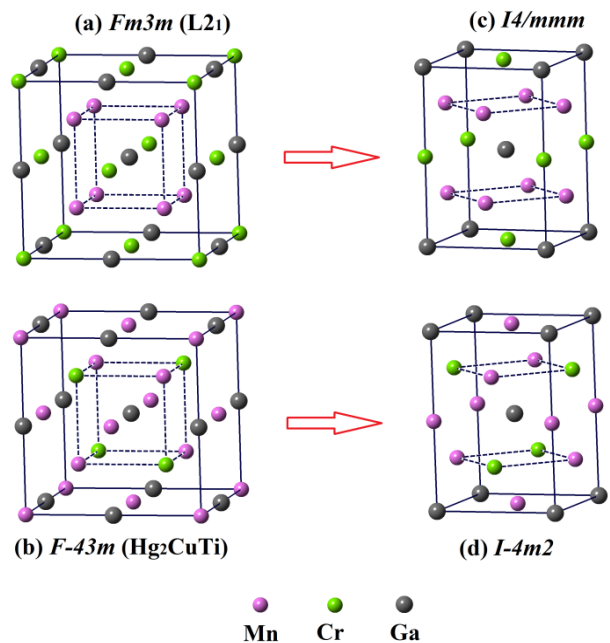

Fig.1 Atomic configuration of Heusler alloys with cubic (austenite phase) and tetragonal structure (martensite phase). (a) $L2_1$ structure (*Fm3m*); (b) $Hg_2CuTi$ structure (*F-43m*); (c) martensite phase of $L2_1$ structure (*I4/mmm*); (d) martensite phase of $Hg_2CuTi$ structure (*I-4m2*).

2. COMPUTATIONAL AND EXPERIMENTAL DETAILS

Before proceeding further, the atomic configuration of Heusler alloys with cubic (austenite phase) and tetragonal structure (martensite phase) need to be addressed. It follows a simple rule that the transition elements with more valence electrons have the priority to occupy those sites nearest to the main-group elements[19-21]. Accordingly, Mn$_2$CrGa is supposed to adopt the so-called L2$_1$ structure shown in Fig.1 (a), where Ga atoms take the *D*(0.5, 0.5, 0.5) sites, Cr atoms take the *B*(0, 0, 0) sites while Mn atoms take the *A*(0.25, 0.25, 0.25) and *C*(0.25, 0.25, 0.75) sites. The martensite corresponding to this structure possesses an *I4/mmm* symmetry as shown in Fig.1 (c). However, there always a certain degree of anti-site disordering in Heusler alloy arose in practice between *A/C* and *B* sites. One extreme situation is that the atoms on *A* or *C* site exchanges with those on B site completely, which is called the Hg$_2$CuTi structure with Mn on *A* and *B* sites, as shown in Fig.1 (b). The corresponding martensite is shown in Fig.1 (d), which possesses an *I-4m2* symmetry. Conversion of the unit-cell from cubic to tetragonal structure is also shown in Fig. 1, which simply obeys that $c_{tet} = a_{cub}$, $a_{tet} = \sqrt{2}a_{cub}$.

The first-principles method used in the present work is based on the density-functional theory.[22] The electronic structure of Mn$_2$CrGa is calculated by using the CASTEP code.[23,24] The interactions between the atomic core and the valence electrons were described by the ultrasoft pseudopotential [25]. The electronic exchange-correlation energy was described with the generalized-gradient approximation (GGA)[26], which provides better agreement with experiment for Heusler alloys compared to local-density approximation.[13,27] The plane-wave basis set cut-off was used as 500 eV for all the cases. The Brillouin zone was sampled by a uniform *k*-point mesh. 120 and 288 *k* points were employed in the irreducible Brillouin zone for cubic austenitic and non-modulated tetragonal martensitic phases, respectively. . The total-energy difference tolerance for the self-consistent field iteration was set at 5 × 10$^{-7}$ eV per atom.

The Mn$_2$CrGa ingots were prepared by arc-melting methods from pure elements. The purity of the raw materials was 99.9% or higher. All the ingots were melted at least four times for homogenization. A very obvious precipitation of Cr from the alloy has been detected by scanning electron microscope (SEM). To prohibit this phase separation, the ingots were then melted and spun into ribbons at 35 m/s, annealed subsequently at 1173 K for 3 days under protection of argon atmosphere. X-ray powder diffraction (XRD) with Cu *K$_\alpha$* radiation was used to verify the crystal structure at room temperature. Thermal analysis was performed using differential scanning calorimetry (DSC). The ramping rate is 10 K min$^{-1}$. Magnetization measurements were carried out in a vibrating-sample magnetometer (VSM) and in a superconducting quantum interference device (SQUID).

## 3. RESULTS AND DISCUSSIONS

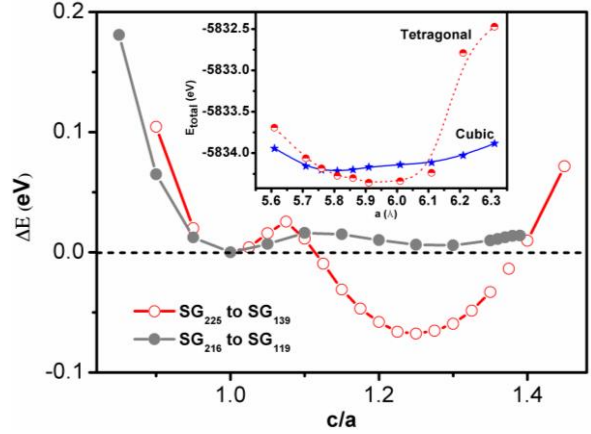

Fig.2. The variation of total energy change as the structure is distorted from L2$_1$ (SG$_{225}$)/Hg$_2$CuTi (SG$_{216}$) cubic to tetragonal type via changing the $c/a$ ratio. The inset shows the $E_{total}$ as a function of lattice constant $a$ for the cubic and tetragonal structures with fixed $c/a$=1.25.

Fig.2 shows the total energy ($E_{total}$) difference between the L2$_1$/Hg$_2$CuTi cubic and non-modulated tetragonal structure, as a function of the distortion along $c$ axis ($c/a$). If a certain tetragonal phase has lower energy than the cubic phase, under the hypothesis of volume conserving, then the compound may have the possibility to show a phase transition effect. In the present system, there are two minima correspond to both L2$_1$ ($c/a$=1, 1.25) and Hg$_2$CuTi ($c/a$=1, 1.3) structures on Fig.1. However, for the highly ordered L2$_1$ structure, the tetragonal phase with $c/a$=1.25 has a much lower energy than that of the cubic phase. On the other hand, for the Hg$_2$CuTi structure, the lowest energy of tetragonal phase has almost the same value with the cubic phase. Therefore, one can come to the conclusion that Mn$_2$CrGa alloy with L2$_1$ structure should trend to transform into a tetragonal structure, which will be invalid if it has an Hg$_2$CuTi structure. This clearly indicates that the atomic ordering has crucial influences on the structure stability of this alloy.

For the L2$_1$ structure, we further evaluate the stability of tetragonal structure on account of the unit cell volume variation with fixed $c/a$=1.25. The inset of Fig.2 gives the volume dependence of the $E_{total}$ for tetragonal and cubic structures. It shows that $E_{total}$ for the tetragonal structure is lower than that of the cubic structure in between $a$=5.75 and 6.1 Å. Since the tetragonal structure usually is the lower temperature phase, larger difference between these two energies would imply its greater stability.

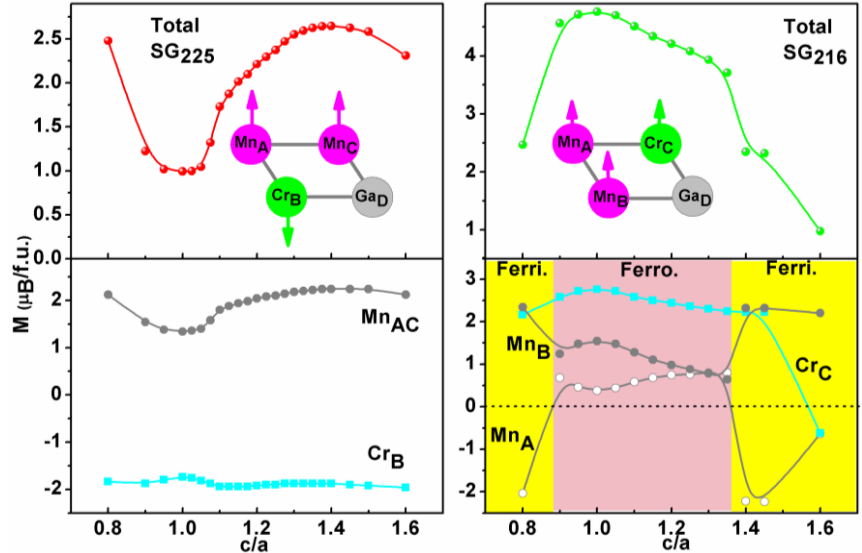

Fig. 3 The evolution of total and atomic magnetic moment along with the structure distortion for L2$_1$.

Fig. 3 shows the evolution of magnetic moment along with the structure distortion. Fig.3 (a) clearly shows that high moment is preferred by the potential structure transition in L2$_1$ structure. In this structure, the Mn atoms on *A&C* sites make the largest contribution to the total magnetic moment ($M_{total}$). While the Cr atoms on *B* sites also possesses a large moment ($M_{Cr}$) which is opposite to that of Mn atoms ($M_{Mn}$), Fig.3 (b). However, the situation in Hg$_2$CuTi-structure is more complex. Fig.3 (c) shows that the $M_{total}$ of Hg$_2$CuTi structure is monotonously decreased by the increase of distortion from cubic. Fig.3 (d) presents the details of the variation of the atomic moments in Hg$_2$CuTi structure. In the *c/a* range of 0.9~1.35, the system has a ferromagnetic structure and the decrease is originated from the decrease of moments on Mn$_B$ and Cr$_C$. Beyond this rang of distortion, ferrimagnetic structures with anti-parallel Mn$_A$ and Mn$_B$/Cr$_C$ is obtained.

There are two points about the magnetic structure of the cubic Hg$_2$CuTi-structure needed to be mentioned. First is that the Cr$_C$ atoms have the largest moment compared to Mn$_A$ and Mn$_B$. This is not like the usual situation in Heusler alloy in which the atoms on B sites should have the largest moment.[] The other point is the unusual ferromagnetic structure in such a Mn and Cr based Heusler alloy. It gives Mn$_2$CrGa a relatively large $M_{total}$ of about 4.8 $\mu_B$, which is rarely seen in Heusler family considering its small valence electrons (23 per unit cell).[28]

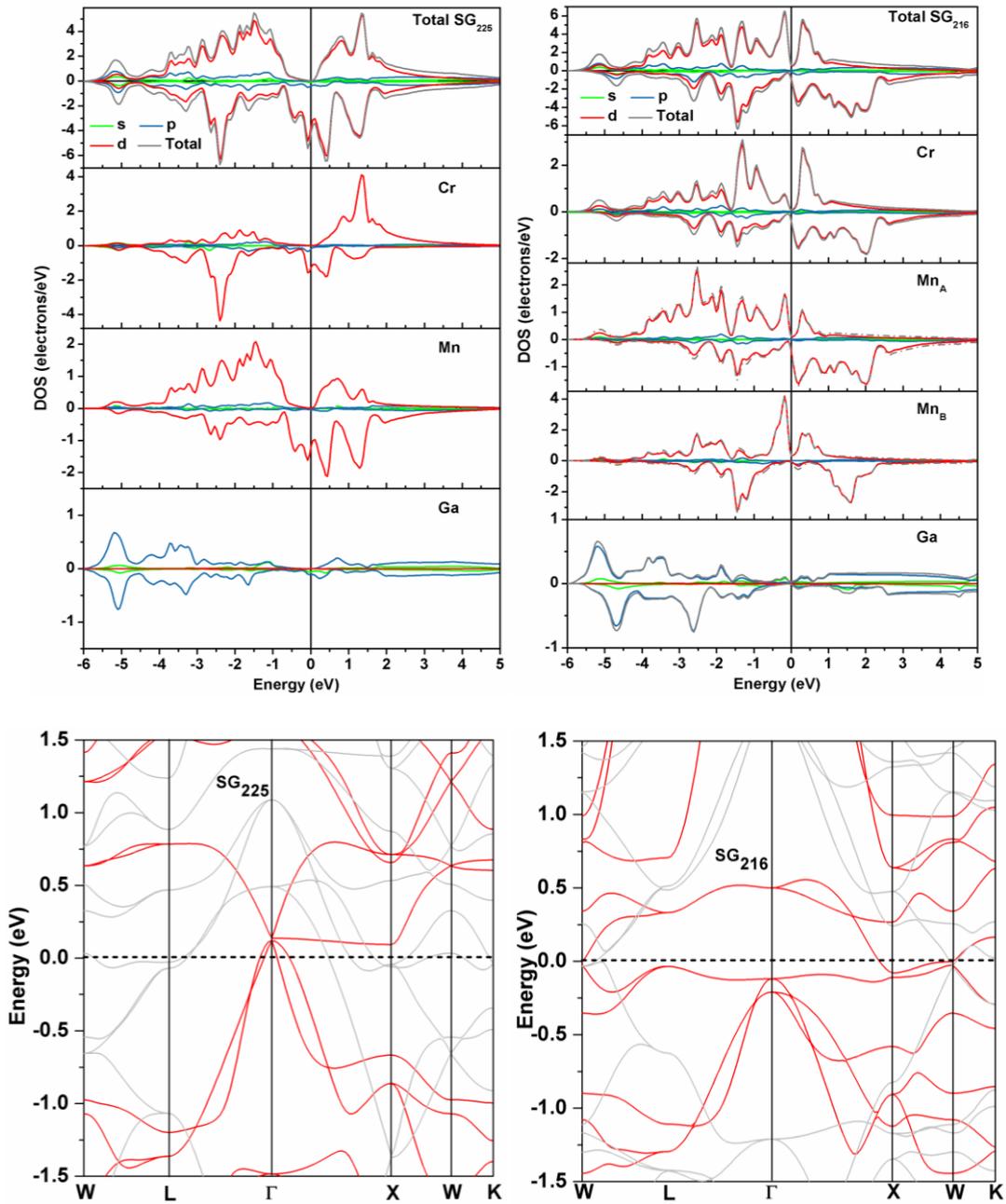

Fig. 4 Spin-polarized total and atom-projected densities of states of Mn$_2$CrGa calculated at the cubic phase (c/a=1) of (a) $L2_1$ (G$_{225}$) and (c) Hg$_2$CuTi (SG$_{216}$). The Fermi energy is considered as zero.

Fig.4 (a) and (b) show the spin-projected total (DOS) and atom-projected (PDOS) densities in the cubic phase (c/a=1) of $L2_1$ (G$_{225}$) and Hg$_2$CuTi (SG$_{216}$), respectively. It shows that the $d$ orbitals make the most important contribution to the DOS near Fermi level. In Fig.4 (a), there exists a pseudo-gap at the Fermi level in the majority (spin-up) DOS of the $L2_1$ structure, which develops into an actual gap at 0.12 eV above the Fermi level with a width of 0.018 eV (Fig.4 (b)). This indicates the covalent bonding characters between the atoms in $Mn_2CrGa$, which is mainly formed by the hybridization between the minority electronic states of Mn and Ga based on their local DOS.[29,30] This is correspond to the nearest neighborship of Mn and Ga, from which point of view a strong interaction between them is expected. In addition, the fully occupied bonding states below the pseudo-gap and the almost empty antibonding states above the pseudogap implies that the covalent bond in the $L2_1$ structure is strong.

Nevertheless, the most important feature in Fig.4 (a) is emerged in the minority (spin-down) part. There is a substantial peak located almost right on the Fermi level, which is also contributed mostly by the Mn atoms. This is a sign of the instability of the $L2_1$ structure, which means that the atoms may tend to show anti-sites disordering in this system. Besides of this, the structure also shows a clear half-metal-like electronic behavior as in previous references. The spin polarization degree (SPD) is calculated to be 99.5% at the Fermi level according to the definition

$$\text{SPD} = \frac{n\uparrow(E_F) - n\downarrow(E_F)}{n\uparrow(E_F) + n\downarrow(E_F)}$$

where $n\uparrow(E_F)$ and $n\downarrow(E_F)$ is the DOS at the Fermi level in the majority (minority).

In Fig.4 (a), the $Hg_2CuTi$ structure presents a pseudo-gaps at the Fermi level in both the majority and minority DOS. From this point of view, this structure is supposed to be more stable than the $L2_1$ structure, i.e. the Mn-Cr atoms disordering is preferred in $Mn_2CrGa$. More specifically, the bonding and anti-bonding majority peaks at around -0.18 and 0.3 eV, respectively, have contributions from Cr, $Mn_A$ and $Mn_B$ $d$ electrons. On the other hand, the minority bonding peaks at -1.4 eV come from all the three magnetic atoms. However, the contributions toward the anti-bonding states in the minority bands near the Fermi levels come mainly from the $d$ electrons of Cr and $Mn_A$. The structures of the bonding DOS, away from the Fermi levels consists of three parts: the $d$-$d$ hybridizations between the Cr and $Mn_A$, the $d$-$d$ hybridizations between the Cr, $Mn_A$ and $Mn_B$ as well as the $p$-$d$ hybridizations between Ga and Cr ($Mn_A$).

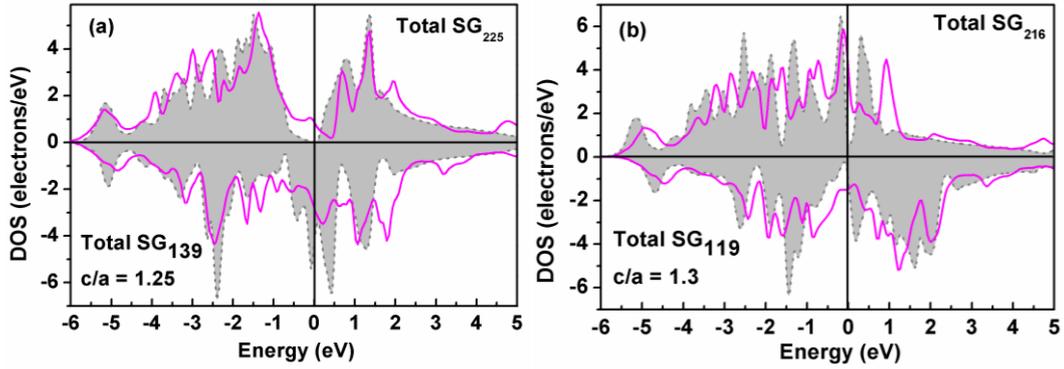

Fig. 5. Spin-polarized DOS of Mn$_2$CrGa calculated at the cubic phase (c/a=1. gray) and martensitic phase (c/a=1.25, 1.3, magenta) for (a) $L2_1$ and (b) Hg$_2$CuTi structure, respectively.

The DOS structures of the cubic phases are severely changed by large tetragonal distortion (c/a>1.25) which breaks the symmetries of the $d$ orbitals. For $L2_1$ structure, the minority spin peaks on the Fermi level is strongly reduced and breaks into multiple peaks. Though the DOS in the majority band at Fermi level is increased slightly, an overall shift of states toward lower energies still make the martensitic phase more stable than the $L2_1$ Cubic phase. This is consistent with the result of Fig.2. On the other hand, the DOS structures of Hg$_2$CuTi structure becomes unstable after the tetragonal distortion. As shown in Fig.5 (b), the original pseudo-gaps in both majority and minority bands are filled. The majority peak at -0.18 eV moves even closer to the Fermi level, and also the whole minority band shifts toward higher energies. These migrations of electronic states all lead to the instability of the tetragonal phase derived from Hg$_2$CuTi structure.

Based on the above theoretical results, we can come to a conclusion that, though the martensitic phase is supposed to existence for a highly ordered $L2_1$ structure, an anti-sites disordering tendency of the Mn-Cr atoms gives rise to a more stable cubic phase than the tetragonal martensitic phase. Therefore, the degree of ordering in the actual Mn$_2$CrGa sample will decide whether the material presents phase transition phenomenon or not.

So far there is no experimental results on this system, therefore, to confirm these theoretical results, we prepared the Mn$_2$CrGa compound by using arc-melting and the followed melting spun method. Fig.6 shows the room temperature XRD pattern of the obtained sample, together with the calculated pattern of the high ordering $L2_1$ structure of the compound. The Mn$_2$CrGa compound possesses a pure cubic Heusler structure, corresponding to the calculated one. One may also noticed that there are two superlattice diffraction peaks on the calculated pattern indexed as (111) and (200), which are the indication of the degree of atomic ordering in the system. However, these two peaks are all absent in the experimental pattern. This implies a highly disordered configuration of the atoms in this compound we prepared. One reason is the nonequilibrium method we used to prepare the sample, which usually will introduce certain degree of disordering in a system. More importantly, the anti-sites disordering of Mn and Cr atoms should be the nature of this system, as the above calculations indicated.

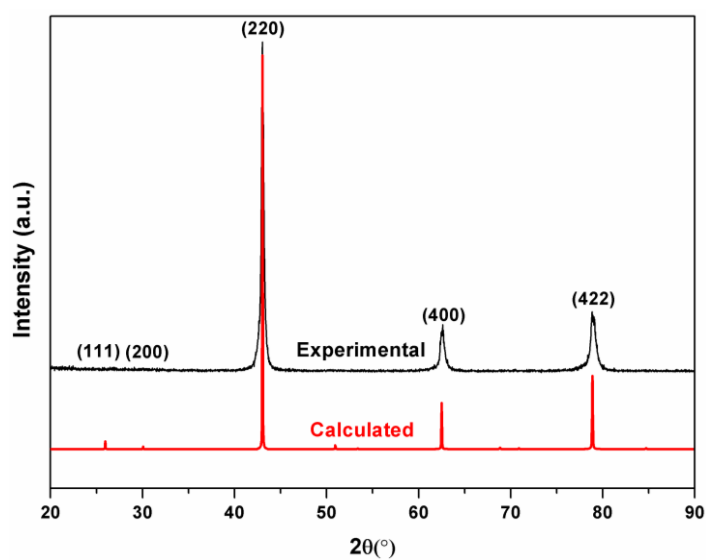

Fig.6 Room temperature XRD pattern of the $Mn_2CrGa$ sample and the calculated one of the high ordering $L2_1$ structure $Mn_2CrGa$ compound.

As for the magnetic behavior, Fig.7 shows the temperature (MT) and field (MH) dependences of the magnetization of the compound. There is no sign of temperature induced martensitic transition in the whole measuring temperature range from 5 K to 1000 K. This is expected, according to our calculation, for this compound with deviated atom atomic configuration from the ideal $L2_1$ structure. Despite the absence of martensitic transition, there still some noticeable features in this system. First is the relatively high Curie temperature ($T_C$) of about 900 K, which is rarely seen in Mn-based Heusler alloys. This indicates a very strong exchange interaction between the magnetic atoms. The other one is the remarkable Hopkinson-like effect appeared on the MT curve, after which the magnetization decreases monotonously along the decreasing temperature. The accepted explanation of the Hopkinson effect is based on the variation of anisotropy against temperature, superparamagnetic-like transition and spin-glass-like behavior. In the present system, considering the possible antiferromagnetic exchange interactions between Mn-Mn, Mn-Cr and Cr-Cr atoms, there should exist an intense competition between them which should be responsible for such phenomenon. This could be further proved by the MH curve at 5 K. The curve is still not saturated at 30 kOe, implying the antiferromagnetic matrix and thus a strong anisotropy of the system. Besides, the sample possesses a moment of 0.15 $\mu_B$ per unit cell under 30 kOe, which is much smaller than our calculated results (seen Fig.2). This is not only due to the disordering of the magnetic atoms, but may also related to a more complicated magnetic structure other than the theoretical collinear one.

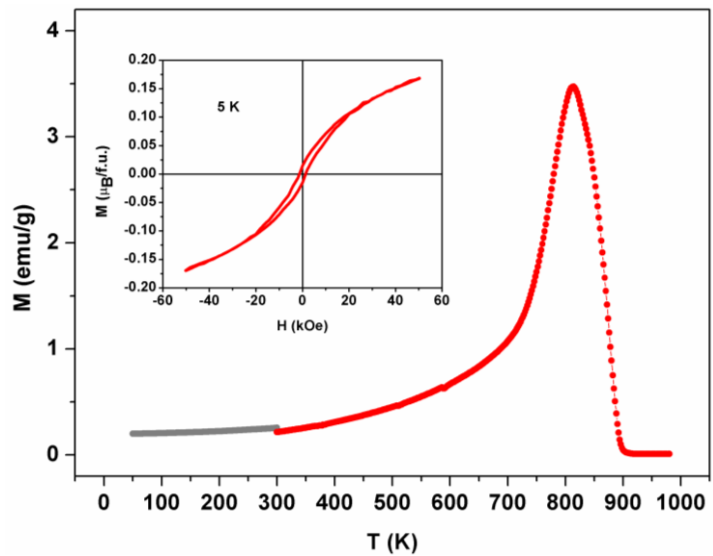

Fig.7 MH curve of the $Mn_2CrGa$ sample under 100 Oe. The inset shows the MH curve at 5 K.

4. **CONCLUSION**

Theoretical investigations on the possibility of martensitic phase transformation in $Mn_2CrGa$ compound with different structures shows that, with Mn atoms on the A/C sites and Cr on B sites, this compound could has a phase transition occurred at certain temperature. On the other hand, anti-sites disordering between Mn and Cr will lead to a clear enhancement of the energy of the martensitic tetragonal phase and suppress the phase transition of the system. Experimental characterizations of the $Mn_2CrGa$ compound have been carried out for the first time to prove the theoretical results. It shows a severe disordering took place in this system, which prohibits the transition from cubic to tetragonal phase as the calculations indicated. The experimental results also imply a strong anisotropy and complicated non-collinear magnetic structure in this system, which will be thoroughly studied later on.

ACKNOWLEDGEMENTS:

This work was supported by the National Natural Science Foundation of China (Grant Nos.51401002 and 51171003).